\documentclass[conference]{IEEEtran}
\IEEEoverridecommandlockouts
\usepackage{cite}
\usepackage{url}
\usepackage{amsmath,amssymb,amsfonts}
\usepackage{algorithmic}
\usepackage{graphicx}
\usepackage{textcomp}
\usepackage{xcolor}
\usepackage{listings}
\usepackage{multirow} 
\usepackage{array}
\usepackage{booktabs}
\usepackage[ruled,vlined]{algorithm2e}
\usepackage{subcaption}

\def\BibTeX{{\rm B\kern-.05em{\sc i\kern-.025em b}\kern-.08em
    T\kern-.1667em\lower.7ex\hbox{E}\kern-.125emX}}

\DeclareRobustCommand*{\IEEEauthorrefmark}[1]{%
	\raisebox{0pt}[0pt][0pt]{\textsuperscript{\footnotesize\ensuremath{#1}}}}

\usepackage[noabbrev,nameinlink]{cleveref}
\Crefformat{figure}{#2Fig.~#1#3}
\Crefmultiformat{figure}{Figs.~#2#1#3}{ and~#2#1#3}{, #2#1#3}{ and~#2#1#3}

\begin{document}

\title{CBCMS: A Compliance Management System for Cross-Border Data Transfer}

\author{
	\IEEEauthorblockN{Zhixian Zhuang\IEEEauthorrefmark{1,2}, Xiaodong Lee\IEEEauthorrefmark{1,3,4}\textsuperscript{,\textsection}, Jiuqi Wei\IEEEauthorrefmark{1,2}, Yufan Fu\IEEEauthorrefmark{1,2}, Aiyao Zhang\IEEEauthorrefmark{1,2}}
	
	\IEEEauthorblockA{\IEEEauthorrefmark{1}Institute of Computing Technology, Chinese Academy of Sciences}
	
	\IEEEauthorblockA{\IEEEauthorrefmark{2}University of Chinese Academy of Sciences}

        \IEEEauthorblockA{\IEEEauthorrefmark{3}Fuxi Institution}

        \IEEEauthorblockA{\IEEEauthorrefmark{4}Center for Internet Governance, Tsinghua University}
	
	\IEEEauthorblockA{\{zhuangzhixian22s, xl, weijiuqi19z, fuyufan20z, zhangaiyao22z\}@ict.ac.cn}
} 
\maketitle

\begingroup\renewcommand\thefootnote{\textsection}
\footnotetext{Xiaodong Lee is the corresponding author.}
\endgroup

\begin{abstract}
Cross-border data transfer is vital for the digital economy by enabling data flow across different countries or regions.
However, ensuring compliance with diverse data protection regulations during the transfer introduces significant complexities. 
Existing solutions either focus on a single legal framework or neglect real-time and concurrent processing demands, resulting in incomplete and inconsistent compliance management. 
To address this issue, we propose Cross-Border Compliance Management System (CBCMS), which not only enables the unified management of data processing policies across multiple jurisdictions to ensure compliance with various legal frameworks involved in cross-border data transfer, but also supports real-time and high-concurrency processing capabilities. 
We design Policy Definition Language (PDL) that supports the unified management of data processing policies, bridging the gap between natural language policies and machine-processable expressions, thereby allowing various legal frameworks to be seamlessly integrated into CBCMS.
We present Compliance Policy Generation Model (CPGM), the core component of CBCMS, which generates compliant data processing policies with high accuracy, achieving up to 25.16\% improvement in F1 score (reaching 97.32\%) compared to rule-based baseline.
CPGM achieves inference time in the order of milliseconds (6 to 13 ms), and keeps low latency even under high-load scenarios, demonstrating high real-time and concurrent performance. 
To our knowledge, CBCMS is the first system to support unified compliance management across jurisdictions while ensuring real-time and concurrent processing capabilities.
\end{abstract}


\section{Introduction}
\subsection{Motivation}
In the digital age, data is a crucial asset that drives the growth and innovation of the digital economy, enabling businesses to optimize operations, make data-driven decisions, and provide personalized services \cite{data1,data2,data3}. 

Cross-border data transfer involves the movement of data from one legal jurisdiction to another, which is essential for global businesses and international operations \cite{data4}. 
For example, a multinational corporation may need to transfer both European customer data and data from California residents to its headquarters in California, United States, for centralized processing. 
However, this process introduces significant challenges in ensuring compliance with stringent data protection laws and regulations \cite{survey1,survey2}, such as the EU General Data Protection Regulation (GDPR) \cite{gdpr} and the California Consumer Privacy Act (CCPA) \cite{ccpa}.
In this scenario, the corporation's European branch must ensure GDPR compliance before transferring data, including obtaining explicit consent and ensuring data minimization.
Simultaneously, once the data arrives in California, the headquarters must adhere to CCPA regulations, such as respecting consumer rights and providing data disclosure.
In this context, ensuring compliance requires aligning data processing practices with the differing legal requirements across jurisdictions to protect data security and privacy, which is a complex task.

In cross-border data transfer scenarios, current solutions primarily focus on two aspects.
On the one hand, some solutions are dedicated to ensuring compliance within specific areas, such as privacy policies, application code, and database storage \cite{policy1,policy2,policy4,policy5,code1,code2,code3,db1,db2}.
However, they are generally confined to single legal framework like GDPR and overlook the diversity of policies required for cross-border transfer, limiting their broader applicability and flexibility. 
Moreover, they also fail to enable data owners to flexibly define their policies.
On the other hand, some solutions aim to enhance security and privacy during data transfer by developing blockchain, encryption or infrastructure based methods\cite{blc2,blc3,blc4,blc5,enc1,enc2,enc3,inf1,inf2}.
Unfortunately, they always encounter performance limitations and deployment challenges, which restrict their adoption to broader scenarios, failing to address the demand for real-time and concurrent data processing during cross-border transfer comprehensively.


Therefore, it is crucial to develop a flexible and comprehensive approach to manage data processing policies and adapt to various legal requirements.
This approach should support the unified management of complex policies, including both compliance policies and those specified by data owners, while ensuring the real-time and concurrent processing capabilities necessary for effective cross-border data transfer.

\subsection{Our Solution}

We propose \textbf{C}ross-\textbf{B}order \textbf{C}ompliance \textbf{M}anagement \textbf{S}ystem (CBCMS) to address compliance issues in cross-border data transfer. 
CBCMS allows flexible configuration and extension of data processing policies. 
To seamlessly integrate data processing policies under various legal frameworks and data owner requirements into CBCMS, we introduce Policy Definition Language (PDL). 
PDL bridges the semantic gap between natural language data processing policies and machine-processable expressions, enabling unified policy management across different jurisdictions. 
Based on PDL, we annotate data and train the Compliance Policy Generation Model (CPGM), the core component of CBCMS, which accurately generates compliant data processing policies according to metadata (e.g., data category and sensitivity level) and the legal jurisdictions (source, target or both) involved. 
By following these generated policies, cross-border compliance can be ensured. 
CPGM's inference time is in the order of milliseconds, and it effectively utilizes multithreading to reduce latency and enhance throughput under high concurrency scenarios. 
This ensures that CBCMS maintains high performance in real-time and concurrent processing conditions.

Our contributions are summarized as follows:

\begin{itemize}
    \item We introduce Policy Definition Language (PDL) (Section \ref{c3}), a schema supporting unified management of data processing policies by bridging the gap between natural language policies and machine-processable expressions. 

    \item We present CBCMS (Section \ref{c4}), a system enabling flexible configuration and extension of data processing policies and ensuring compliance across jurisdictions. 
    
    \item We develop CPGM (Section \ref{c5}), a model generating compliant data processing policies accurately and promptly.

    \item We implement and evaluate a prototype system (Section \ref{c6}). 
    CPGM improves F1 score by up to 25.16\% compared to the rule-based baseline, reaching 97.32\%, with an inference time of 6 to 13 ms, maintaining low latency under high-load scenarios. 
    To our knowledge, CBCMS is the first system to unify compliance management across jurisdictions while ensuring real-time and concurrent processing capabilities.

\end{itemize}


\section{Preliminary and Related Work} \label{c2}
\subsection{Preliminary}
\subsubsection{Compliance Requirements}
Increasingly stringent compliance requirements pose challenges for global data governance \cite{survey1,survey2}.
For instance, the General Data Protection Regulation (GDPR) \cite{gdpr}, the California Consumer Privacy Act (CCPA) \cite{ccpa}, and China's Personal Information Protection Law (PIPL) \cite{pipl} impose strict data protection standards on cross-border data transfer. 
In this paper, we develop CBCMS, which are designed to demonstrate our approach's advancements in handling cross-border data transfer under various legal frameworks, including GDPR, CCPA, and PIPL.

\subsubsection{Service Model}
\begin{figure}[tp]
\centering
\includegraphics[width=0.95\columnwidth]{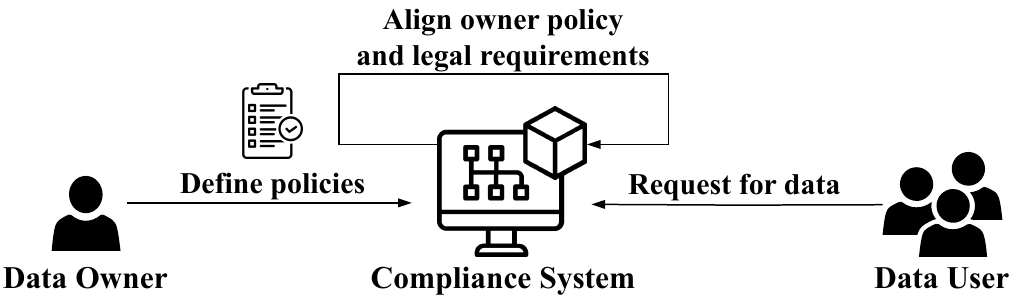}
\caption{CBCMS service model.}
\label{fig:service_model}
\vspace{-10pt}
\end{figure}

Fig. \ref{fig:service_model} shows our service model, which involves data owners, data users, and the compliance system.
The roles and interactions are as follows:

\begin{itemize}
    \item \textbf{Data Owners}: Entities or individuals owning data and are responsible for defining data processing policies. They expect flexibility in configuring these policies to protect their data and ensure compliance with legal requirements.
    
    \item \textbf{Data Users}: Entities or individuals who process the data according to the policies defined by the data owners. They must adhere strictly to these policies to avoid unauthorized data usage and potential legal issues.
    
    \item \textbf{Compliance System}: Mediates the interaction between data owners and data users, ensuring that data processing practices comply with both the policies set by data owners and the legal requirements of the jurisdictions involved.
\end{itemize}

We emphasize that the compliance system is designed to mitigate unintentional non-compliance due to a lack of relevant knowledge among participants of cross-border transfer, not to prevent deliberate violation.

\subsection{Related Work}
\subsubsection{Data Compliance}
With increasing compliance requirements, many studies have focused on data compliance.

\textbf{Privacy policy compliance checking}: Natural language processing (NLP) methods are used to identify GDPR-related information in privacy policies \cite{policy1,policy2,policy4} to check compliance. 
A UML dual-layer representation method is presented to formalize GDPR concepts \cite{policy5}. 
However, these approaches mainly focus on GDPR, neglecting cross-border legal frameworks, failing to meet various compliance requirements.

\textbf{Code compliance checking}: systems are presented for web development frameworks to enforce GDPR compliance \cite{code1,code2}. 
Meanwhile, an automated method evaluating GDPR compliance in cross-border data transfer in Android applications is proposed \cite{code3}. 
These studies face challenges with integration complexity and performance overhead, mainly targeting GDPR and lacking support for cross-border laws.

\textbf{Database compliance checking}: Tools like GDPRizer \cite{db1} and ODLAW \cite{db2} automate the extraction of GDPR-compliant data from legacy databases.
Nevertheless, they primarily focus on GDPR, limiting their extension to other legal frameworks and application scenarios.

\subsubsection{Data Transfer}
Many studies have addressed issues in data transfer.

\textbf{Blockchain-based approaches}: 
Blockchain significantly enhances the security and privacy of data transfer. 
Many blockchain-based approaches are proposed to ensure secure data sharing and transfer \cite{blc2,blc3,blc4,blc5}. 
Despite their benefits, these approaches overlook compliance concerns and demonstrate restricted efficacy, hindering widespread adoption.

\textbf{Encryption-based approaches}: Encryption-based approaches are proposed to ensure secure data transfer in vehicular networks \cite{enc1,enc2} or edge computing environments \cite{enc3}. 
However, these approaches lack support for legal compliance and are limited to specific application scenarios.

\textbf{Infrastructure-based approaches}: Infrastructure-based approaches offer macro solutions for data transfer, such as DATA-EX \cite{inf1} and DIA \cite{inf2}. 
However, they overlook compliance issues and exhibit poor performance, making them unsuitable for real-time and concurrent processing.

\section{Policy Definition Language} \label{c3}
In this section, we introduce our design of \textbf{P}olicy \textbf{D}efinition \textbf{L}anguage (PDL), which describes natural language data processing policies in machine-processable formats, playing a crucial role as the foundation of dataset annotation and CPGM model construction. 
PDL enables structured descriptions of policies from legal mandates or owner preferences, supporting CBCMS's unified policy management.
We also devise a pipeline to map natural language policies into PDL formats with no loss of semantic information.

\begin{figure}[tp]
\centering
\includegraphics[width=0.9\columnwidth]{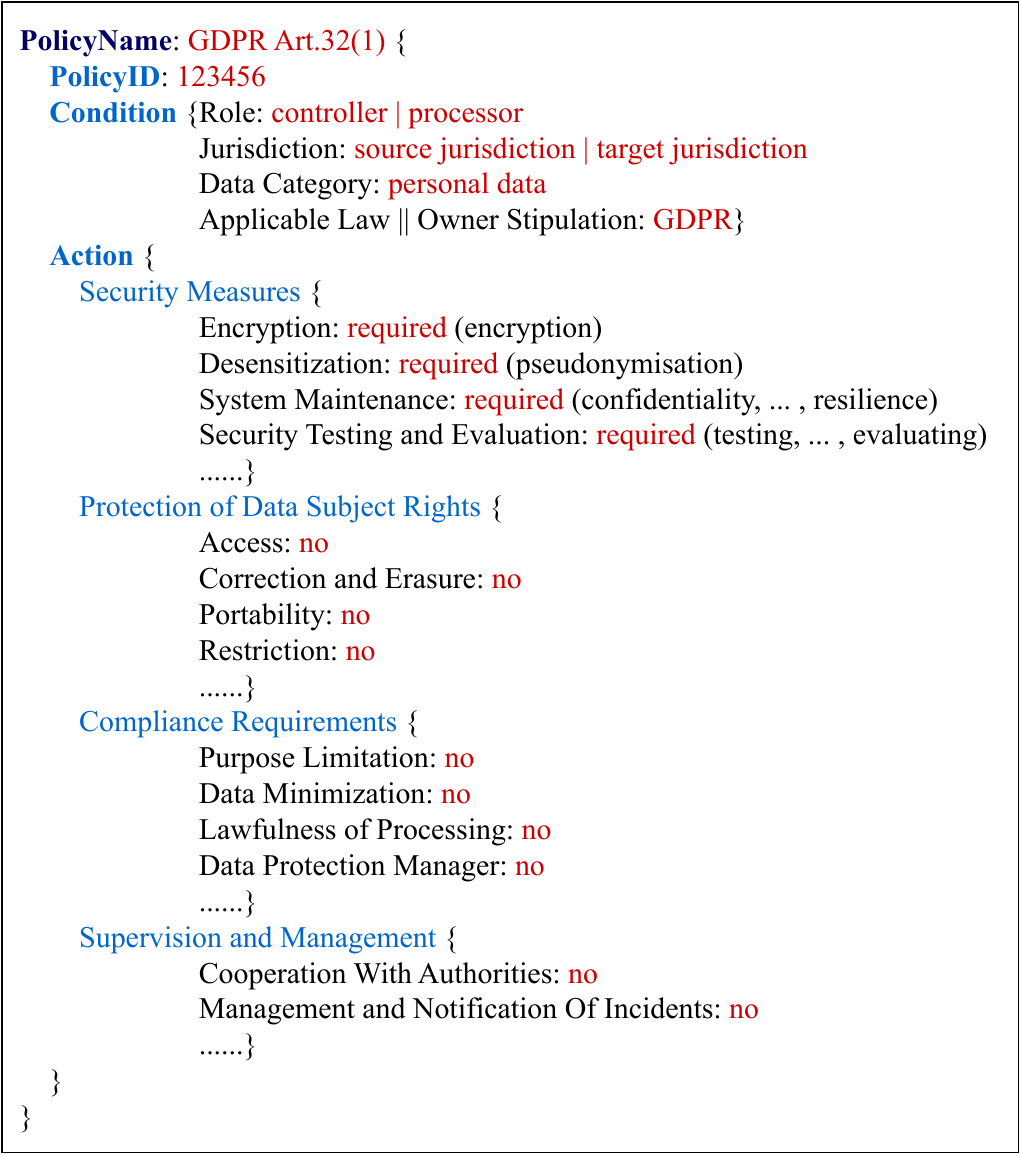}
\caption{The structure and an instance of PDL grammar.}
\label{fig:pdl}
\vspace{-5pt}
\end{figure}

\subsection{PDL Grammar} \label{3.1}
The creation of PDL is centered around its grammar, which defines data processing policies in a format directly interpretable by machines.
Unlike humans, who can parse nuanced natural language, machines require clearly structured information for processing. 
Thus, we design a structured grammatical framework for PDL, as shown in Fig. \ref{fig:pdl}.

\emph{PolicyName} and \emph{PolicyID} identify the name and unique identifier of a data processing policy, respectively. 

\emph{Condition} specifies the prerequisites necessary for implementing a data processing policy.
This includes:

\begin{itemize}
\item \emph{Role}: the entity designated by the policy to take responsibility, such as a ``controller'' or ``processor''.
\item \emph{Jurisdiction}: the jurisdiction(s) to which the policy applies, encompassing the source jurisdiction, the target jurisdiction, or both in cross-border data transfer.
\item \emph{Data Category}: the category of data that the policy governs.
Certain data category may require special treatment.
\item \emph{Applicable Law} or \emph{Owner Stipulation}: the specific legislation or the data owner's stipulations that mandate this policy, such as ``GDPR'' or an internal corporate policy.
\end{itemize}

\emph{Action} outlines the measures required once the \emph{Condition} criteria are satisfied.
This includes four principal categories:

\begin{itemize}
\item \emph{Security Measures}: measures such as data encryption, desensitization, system maintenance and security testing or evaluation.
\item \emph{Protection of Data Subject Rights}: measures to ensure the rights of data subjects, such as access, correction and erasure, data portability and restriction of processing.
\item \emph{Compliance Requirements}: regulations such as purpose limitation, data minimization, lawfulness of processing and the appointment of data protection managers.
\item \emph{Supervision and Management}: cooperation with regulatory agencies and procedures for managing emergency incidents like data breaches.
\end{itemize}


\subsection{PDL Mapping Pipeline} \label{3.2}
The PDL mapping pipeline maps policy texts into machine-processable PDL formats, as shown in Algorithm \ref{a1}.

\begin{algorithm}[tp]
\small
    \caption{PDL Mapping Pipeline}\label{alg:pdl_mapping}
    \label{a1}
    \LinesNumbered 
    \SetNlSty{textbf}{\scriptsize}{}
    \SetAlgoNlRelativeSize{-1} 
    \SetCommentSty{mycommfont} 
    
    \SetKwInput{KwIn}{Input}
    \SetKwInput{KwOut}{Output}
    
    \newcommand\mycommfont[1]{\footnotesize\ttfamily{#1}}
    
    \KwIn{Input text: $L$, Policy Definition Language (PDL) structure: $P$}
    \KwOut{Mapped PDL policies: $M$}
    
    $T \leftarrow$ Tokenize($L$) \tcp{Tokenization}
    $Lemmas \leftarrow$ Lemmatize($T$) \tcp{Lemmatization}
    $Cleaned \leftarrow$ RemoveStopWords($Lemmas$) \tcp{Stop Word Removal}
    $E \leftarrow$ NER($Cleaned$) \tcp{Entity Recognition}
    $R \leftarrow$ DependencyParsing($E$) \tcp{Relation Extraction}
    
    $M \leftarrow \{\}$ \tcp{Initialize mapped PDL policies}
    \For{each entity $e \in E$}{
        $MappedField \leftarrow$ MapEntityToPDLCondition($e, P$)\\
        Add $MappedField$ to $M$
       
        \For{each relation $r \in R$}{
            $MappedAction \leftarrow$ MapEntityAndRelationToPDLAction($e, r, P$)\\
            Add $MappedAction$ to $M$
        } 
    }
    
    \For{each policy $m \in M$}{
        $IsValid \leftarrow$ ValidatePolicy($m$) \tcp{PDL validation}
        \If{$IsValid == false$}{
            $m \leftarrow$ AdjustPolicy($m$)
        }
    }
    \Return $M$
\end{algorithm}

Firstly, the pipeline tokenizes the input text $L$ into individual words and sentences $T$ (line 1), making complex information accessible for automated processing.

Later, the pipeline lemmatizes $L$ into the base form $Lemmas$ (line 2) and removes the stop words to get the cleaned tokens $Cleaned$ (line 3), concentrating on substantive contents critical for policy formulation.

Next, the pipeline focuses on the recognition and analysis of entities.
It extracts significant entities $E$ related to data processing through Named Entity Recognition (NER) (line 4), ensuring that all pertinent entities are accurately identified.
Following NER, dependency parsing is employed to get the analyzed syntactic structure $R$, identifying relationships between entities and their modifiers (line 5).
This helps understand the context associated with each entity and align relevant information into a structured format for detailed analysis.

Then, the pipeline maps the entities and corresponding relations to specific fields, including \emph{Condition} (lines 7-9) and \emph{Action} (lines 10-12) to generate PDL structure $M$.

Finally, each PDL $m$ is validated to ensure compliance and accurate communication of owner intentions (lines 13-16).


\subsection{Application Example: GDPR Art.32(1)} \label{3.3}
\begin{figure}[tp]
\centering
\includegraphics[width=\columnwidth]{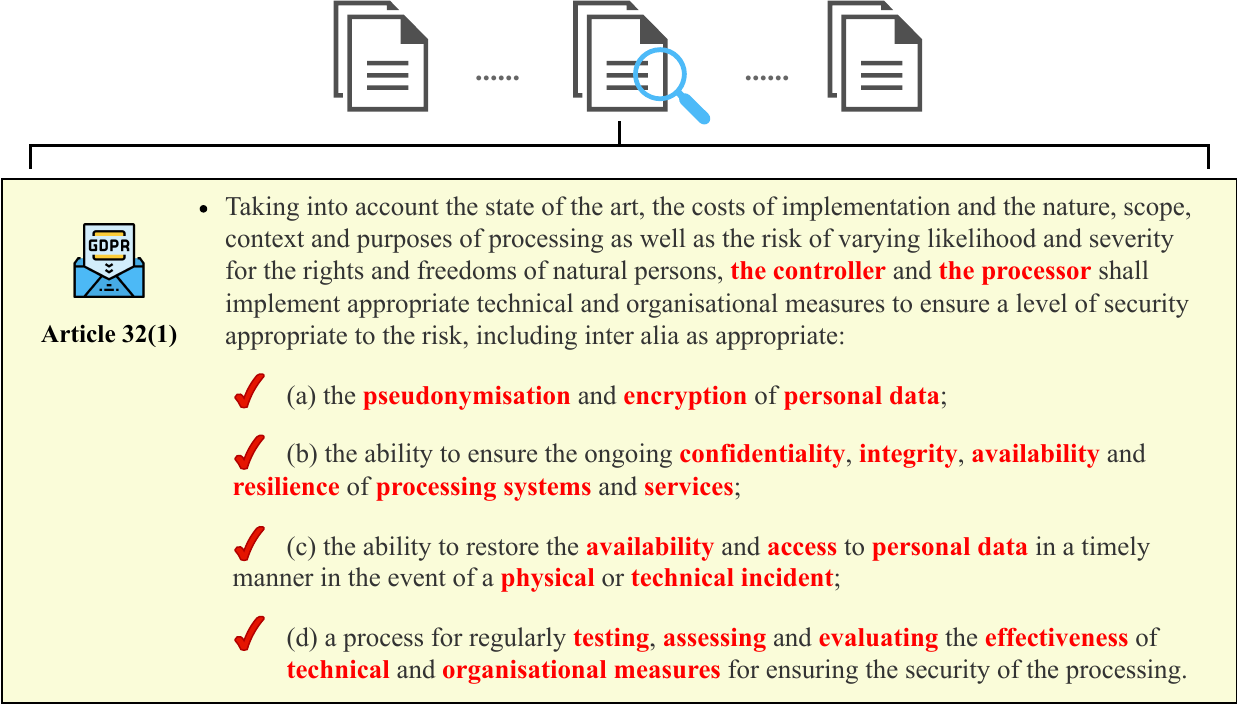}
\caption{Clauses of GDPR Art.32(1).}
\label{fig:gdpr32}
\vspace{-10pt}
\end{figure}

Here we present an example illustrating the practical application of PDL, based on GDPR Art.32(1), which focuses on the security measures data controllers and processors are required to implement, as depicted in Fig. \ref{fig:gdpr32}.

\textbf{Step1.}
The pipeline tokenizes the policy text, performs lemmatization, and removes the stop words. 

\textbf{Step2.}
The pipeline identifies key entities relevant to data processing utilizing NER, as shown in Table \ref{tab:gdpr_entities}.
Next, it parses the dependency of these entities, as illustrated in Table \ref{tab:gdpr_dependencies}. 

\textbf{Step3.}
The pipeline maps the recognized entities to the corresponding fields of the final PDL structure based on the dependency relations.

\begin{itemize}
    \item \emph{Role Entity}: Included in the \emph{Condition - Role} field, specifying who is responsible for implementing the policy.
    \item \emph{Legal Entity}: Specified in the \emph{Condition - Applicable Law} field, providing the legal basis for the policy.
    \item \emph{Data-related Entity}: \emph{Data Category} is specified in the \emph{Condition - Data Category} field, and \emph{Data Media} is used to determine the mapping rules of the \emph{Action} field.
    \item \emph{Constraint Entity}: Specified in the \emph{Condition - Jurisdiction} field, specifying where the policy is effective.
    \item \emph{Action Entity}: Mapped to the \emph{Action - Security Measures} field in the PDL.
\end{itemize}

The final PDL follows the structure defined in Fig. \ref{fig:pdl}.

\begin{table}[t]
\vspace{3pt}
\caption{Named Entity Recognition with GDPR Art.32(1)}
\begin{center}
\renewcommand{\arraystretch}{1.5}
\begin{tabular}{|c|m{5.5cm}|}
\hline
\textbf{Entity Type} & \multicolumn{1}{c|}{\textbf{Details}} \\ \hline
Role Entity & 
Role: controller, processor \\ \hline
Legal Entity & 
Law: GDPR\newline Clause: Article 32(1) \\ \hline
Data-related Entity &
Data Category: personal data\newline Data Media: processing systems, services \\ \hline
Constraint Entity & 
Jurisdiction: source, target \\ \hline
Action Entity & 
Action: pseudonymisation, encryption, confidentiality, integrity, availability, resilience, access, physical, technical, incident, testing, assessing, evaluating, effectiveness, technical measures, organisational measures \\ \hline
\end{tabular}
\label{tab:gdpr_entities}
\vspace{-5pt}
\end{center}
\end{table}

\begin{table}[t]
\caption{Dependency Parsing with GDPR Art.32(1)}
\begin{center}
\renewcommand{\arraystretch}{1.5}
\begin{tabular}{|>{\centering\arraybackslash}p{4.1cm}|>{\centering\arraybackslash}p{3.6cm}|}
\hline
\textbf{Relation} & \textbf{Modifier} \\ \hline
pseudonymisation, encryption & personal data \\ \hline
confidentiality, integrity, availability, resilience & processing systems and services \\ \hline
availability, access & personal data \\ \hline
physical, technical & incident \\ \hline
testing, assessing, evaluating, effectiveness & technical and organisational measures \\ \hline
\end{tabular}
\label{tab:gdpr_dependencies}
\end{center}
\vspace{-10pt}
\end{table}

\section{Cross-Border Compliance Management System} \label{c4}
In this section, we introduce our design of \textbf{C}ross-\textbf{B}order \textbf{C}ompliance \textbf{M}anagement \textbf{S}ystem (CBCMS), which addresses compliance issues in cross-border data transfer by uniformly managing data processing policies and generating compliant policies based on metadata (e.g., data category and sensitivity level) and jurisdictions (source, target or both) involved. 

\begin{figure*}[htbp]
\centering
\includegraphics[width=0.87\textwidth]{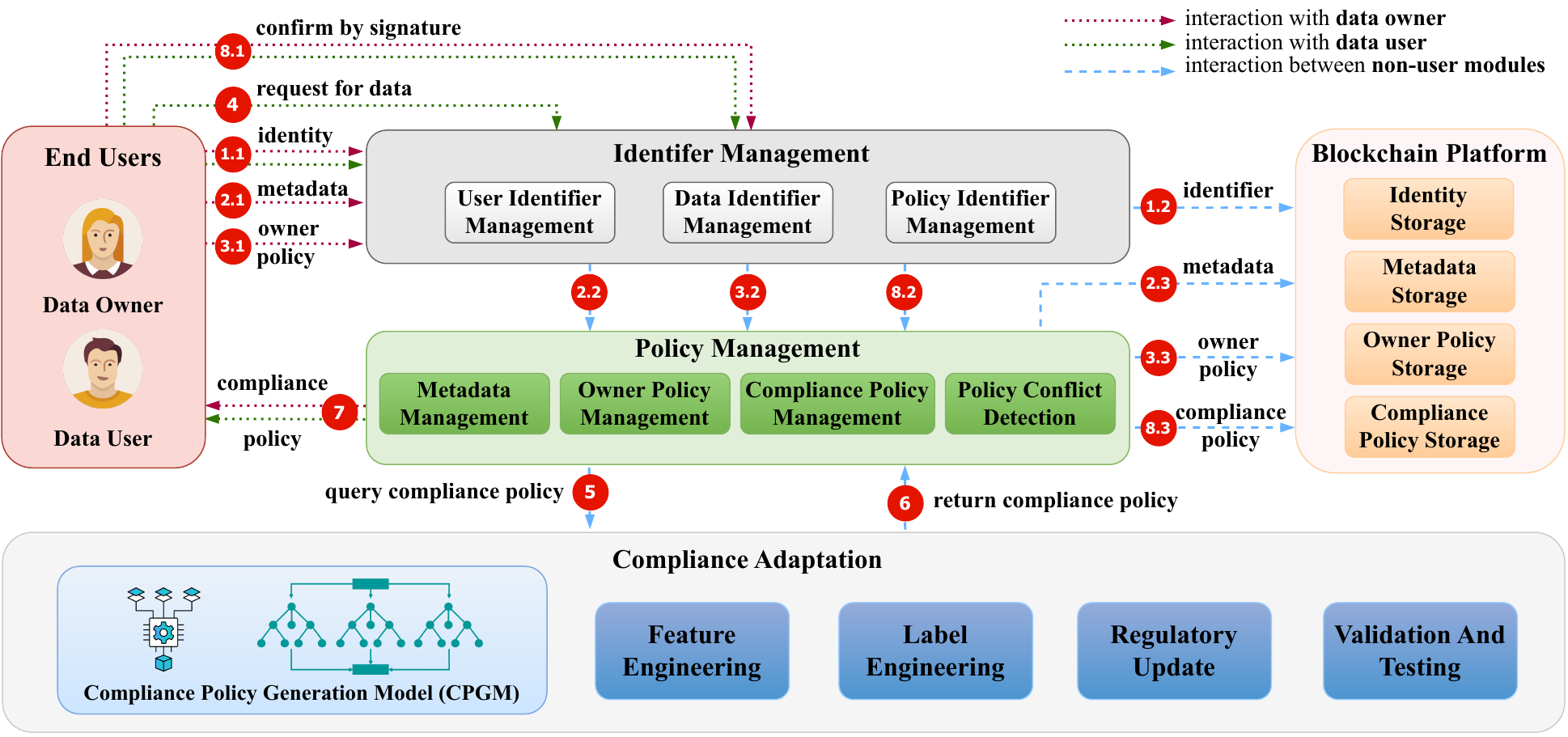}
\caption{System architecture of CBCMS.}
\label{fig:system_architecture}
\vspace{-5pt}
\end{figure*}

\subsection{System Components} \label{4.1}
As shown in Fig. \ref{fig:system_architecture}, CBCMS consists of four main entities: \emph{Identifier Management}, \emph{Policy Management}, \emph{Compliance Adaptation}, and \emph{Blockchain Platform}.
We refer to the users of CBCMS as \emph{End Users}, which represent individuals, organizations, or enterprises that interact with CBCMS.
End users can be data owners or data users.

\subsubsection{Identifier Management} provides globally unique identifiers for users, metadata (e.g., data category and sensitivity level), and policies in CBCMS. We use Universally Unique Identifier (UUID) \cite{uuid} to ensure uniqueness of the identifier.

\textbf{User Identifier Management} provides an identifier for each end user registered in CBCMS, generates a public-private key pair and stores the public key on \emph{Blockchain Platform}.

\textbf{Data Identifier Management} provides an identifier for each piece of metadata registered in CBCMS.

\textbf{Policy Identifier Management} provides an identifier for each data processing policy configured in CBCMS.

\subsubsection{Policy Management} interacts with \emph{Compliance Adaptation} and \emph{Blockchain Platform} to ensure the compliance and transparency of data processing policies.

\textbf{Metadata Management} registers metadata (e.g., data category and sensitivity level), and writes corresponding operation records to \emph{Blockchain Platform}.

\textbf{Owner Policy Management} supports data owners to configure data processing policies in PDL format, and allows flexible addition, modification, or deletion of policies.
It also writes the operation records to \emph{Blockchain Platform}.

\textbf{Compliance Policy Management} receives and stores compliant data processing policies from \emph{Compliance Adaptation}.

\textbf{Policy Conflict Detection} compares owner policies and compliance policies, discovers conflicts between them, and notifies data owners to modify the corresponding data processing policies to meet legal requirements.

\subsubsection{Compliance Adaptation} interacts with \emph{Policy Management} to ensure compliance with various legal requirements.

\textbf{Compliance Policy Generation Model} (CPGM) generates compliant data processing policies based on different jurisdictions and metadata (e.g., data category and sensitivity level), ensuring that all policies meet current legal standards.

\textbf{Feature Engineering} extracts and transforms data features, making them normalized for model training and inference.

\textbf{Label Engineering} converts policies into structured labels, ensures that the labels accurately represent compliance requirements across different jurisdictions.

\textbf{Regulatory Update} keeps CPGM up-to-date with legal changes by tracking and integrating new legal regulations.

\textbf{Validation and Testing} ensures the accuracy and reliability of CPGM through rigorous testing and validation processes. 

When owner policies conflict with compliance requirements, the policies generated by \emph{Compliance Adaptation} take precedence, ensuring compliance while maintaining flexibility for data owners.

\subsubsection{Blockchain Platform} ensures that the entire data processing lifecycle is transparent and traceable.
It interacts with \emph{Policy Management} and includes four functions:

\textbf{Identity Storage} stores the identity information of the end user, including its unique identifier and public key.

\textbf{Metadata Storage} stores the metadata and corresponding operation records.

\textbf{Owner Policy Storage} stores the owner policies and corresponding operation records.

\textbf{Compliance Policy Storage} stores compliance policies and corresponding operation records.

\begin{figure*}[tp]
\centering
\includegraphics[width=0.8\textwidth]{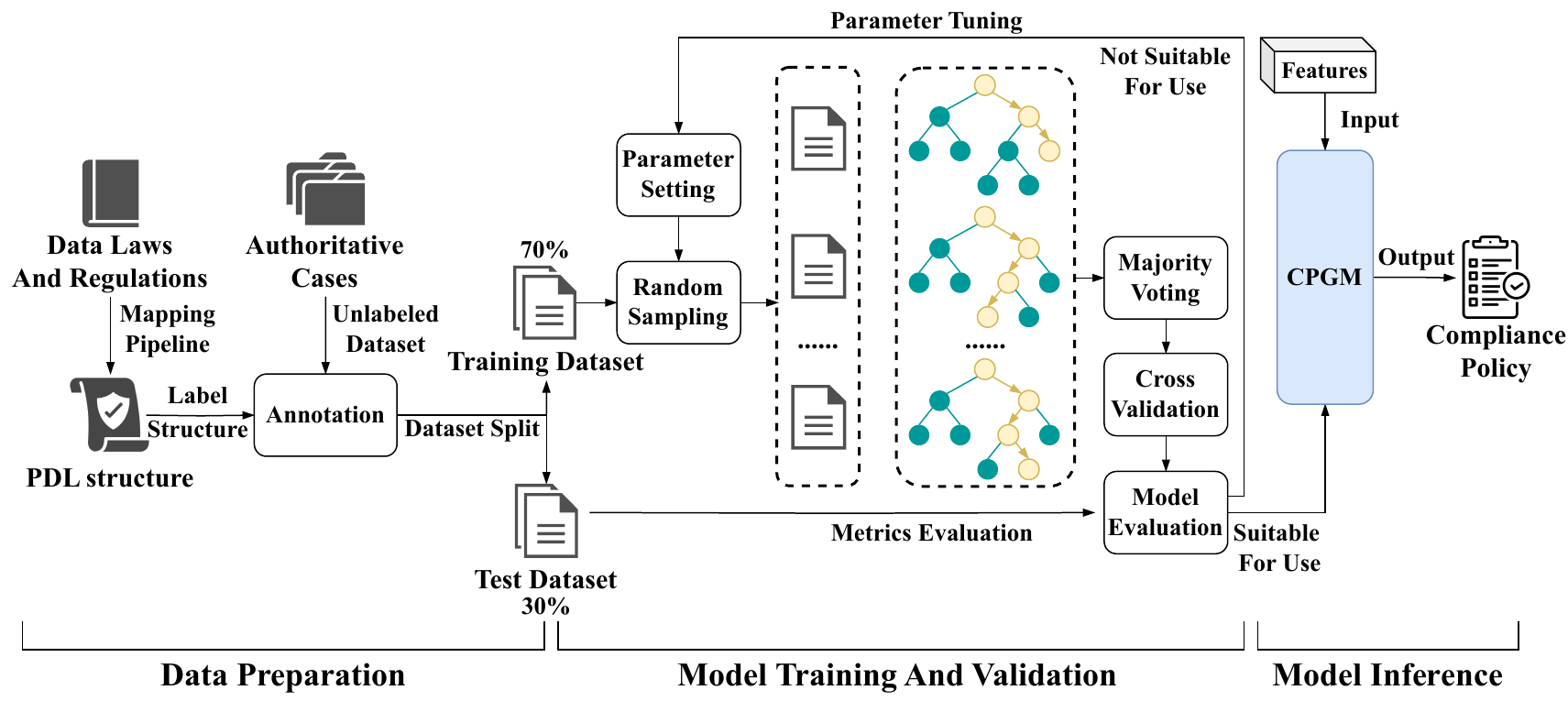}
\caption{Process of CPGM.}
\label{fig:model_process}
\vspace{-5pt}
\end{figure*}

\subsection{Interaction Flow} \label{4.2}
The interaction flow of CBCMS is divided into eight major steps, shown in Fig. \ref{fig:system_architecture}.

\begin{itemize}
    \item \textbf{Step 1.} End users register their identities and get user identifiers in \emph{Identifier Management} (Step 1.1), with their public keys stored on \emph{Blockchain Platform} (Step 1.2).

    \item \textbf{Step 2.} Data owners get data identifiers from \emph{Identifier Management} (Step 2.1).
    Then, \emph{Policy Management} registers the metadata (Step 2.2).
    The operation records are recorded on \emph{Blockchain Platform} (Step 2.3).

    \item \textbf{Step 3.} Data owners get policy identifiers from \emph{Identifier Management} (Step 3.1).
    \emph{Policy Management} (Step 3.2) registers the policies, which are then mapped into PDL formats and stored on \emph{Blockchain Platform} (Step 3.3).

    \item \textbf{Step 4.} Data users initiate data transfer requests in CBCMS, specifying the source and target jurisdictions.

    \item \textbf{Step 5.} The system queries \emph{Compliance Adaptation} for the compliance policies.

    \item \textbf{Step 6.} \emph{Compliance Adaptation} generates compliance  policies and returns them to \emph{Policy Management}.

    \item \textbf{Step 7.} \emph{Policy Management} checks for conflicts between owner and compliance policies. 
    In case of conflict, compliance policies take precedence. 
    The policies are then sent to both data owners and users for confirmation.

    \item \textbf{Step 8.} The end users confirm the data processing policies by digital signatures (Step 8.1).
    \emph{Identifier Management} generates policy identifiers and \emph{Policy Management} registers the policies (Step 8.2).
    Finally, the confirmed data processing policies are mapped into PDL formats and stored on \emph{Blockchain Platform} (Step 8.3).
\end{itemize}

\section{Compliance Policy Generation Model} \label{c5}
We construct a \textbf{C}ompliance \textbf{P}olicy \textbf{G}eneration \textbf{M}odel (CPGM) in \emph{Compliance Adaptation} of CBCMS, which can generate compliance policies for different jurisdictions based on different metadata (e.g., data category and sensitivity level). The process of CPGM is shown in Fig. \ref{fig:model_process}.

\subsection{Data Preparation} \label{5.1}
We use GDPR, CCPA, and PIPL as examples to construct CPGM. 
As shown in Fig. \ref{fig:model_process}, we generate the label structure based on PDL formats of these regulations via the PDL mapping pipeline introduced in Section \ref{3.2}.
The \emph{Action} field of each PDL policy is encoded into CPGM labels using multi-label binary encoding, where 1 indicates an action is required and 0 indicates it is not required.

Next, we collect cases of various cross-border data transfer scenarios and manually annotate the dataset.
To ensure the high quality of our dataset, we only select cases from official government and authoritative organizations, such as reports and guidelines issued by the European Commission, the Federal Trade Commission (FTC) of the United States, and the Cyberspace Administration of China.
Additional cases are sourced from the Privacy Enforcement Casebook \cite{iapp} of the International Association of Privacy Professionals (IAPP), which includes a large number of global privacy cases.

The annotated dataset contains 4923 entries, each with data category, sensitivity level, source and target jurisdiction as features, and the multi-label binary encoded labels as ground truth, as shown in Table \ref{tab:dataset_entry}.
Among these entries, the dataset includes 1928 entries for GDPR, 1598 entries for CCPA, and 1397 entries for PIPL.
The dataset is then divided into a training set (70\%) and a test set (30\%) to ensure effective evaluation and verification of CPGM after training.

\begin{table}[t]
\vspace{4pt}
\caption{Example of an Annotated Dataset Entry}
\begin{center}
\renewcommand{\arraystretch}{1.5}
\begin{tabular}{|>{\centering\arraybackslash}p{1.2cm}|>{\centering\arraybackslash}p{1.2cm}|>{\centering\arraybackslash}p{1.3cm}|>{\centering\arraybackslash}p{1.3cm}|>{\centering\arraybackslash}p{1.6cm}|}
\hline
\textbf{Data Category} & \textbf{Sensitivity Level} & \textbf{Source Jurisdiction} & \textbf{Target Jurisdiction} & \textbf{Encoded Labels} \\ \hline
Personal Data          & Low (level = 1)                      & GDPR (EU)                        & CCPA (CA, USA)                      & [0, 1, ... 0, 1, 0, 1, 1]              \\ \hline
\end{tabular}
\label{tab:dataset_entry}
\end{center}
\vspace{-15pt}
\end{table}

\subsection{Model Training And Validation} \label{5.2}
 To construct CPGM, we choose the Random Forest (RF) \cite{rf} algorithm, which offers unique advantages for addressing compliance issues in cross-border data transfer:

\textbf{Interpretability}: 
    Random Forest offers feature importance \cite{rf_int}, essential for interpreting the model's decisions in compliance scenarios. 
    In contrast, models like deep neural networks and large language models (LLMs) have opaque decision processes, making interpretation difficult. 
    Moreover, LLMs can produce hallucinations and inconsistent predictions \cite{survey3}, making them less reliable for compliance management.

\textbf{High Real-time Performance}:
    Random Forest can quickly generate predictions due to its ensemble of low-complexity decision trees \cite{tree}, ensuring fast overall speeds. 
    Conversely, models like deep neural networks and LLMs have time-consuming training and prediction processes, struggling to meet real-time performance demands.

\textbf{Handling of Imbalanced Datasets}:
    Random Forest effectively handles imbalanced datasets, maintaining good performance despite uneven sample sizes.

Random forest works by creating multiple decision trees from random subsets of the training dataset, and aggregating their predictions through majority voting, shown in Fig. \ref{fig:model_process}.

Considering that our dataset is unbalanced, we use precision, recall, and F1 score to comprehensively evaluate the performance of CPGM. 
These metrics provide a balanced view of the model's ability to make accurate predictions (\( \text{Precision} = \frac{TP}{TP + FP} \), where \(TP\) is true positives and \(FP\) is false positives), identify relevant instances (\( \text{Recall} = \frac{TP}{TP + FN} \), where \(FN\) is false negatives), and balance both aspects (\( \text{F1 score} = 2 \cdot \frac{\text{Precision} \cdot \text{Recall}}{\text{Precision} + \text{Recall}} \)).

We use 5-fold cross-validation during training to evaluate CPGM's generalization ability, avoid overfitting, and ensure robustness on the imbalanced dataset.
Grid search is employed for parameter tuning, with models evaluated primarily on the F1 score.
The parameter settings with the highest F1 score are selected, as shown in Table \ref{tab:parameters}.

\begin{table}[t]
\vspace{4pt}
\caption{Grid Search Parameter Settings and Final Values}
\begin{center}
\renewcommand{\arraystretch}{1.5}
\begin{tabular}{|>{\centering\arraybackslash}p{3.1cm}|>{\centering\arraybackslash}p{2.1cm}|>{\centering\arraybackslash}p{2.1cm}|}
\hline
\textbf{Parameter} & \textbf{Range} & \textbf{Final Value} \\ \hline
Number of trees & [50, 100, 150] & 100 \\ \hline
Maximum depth & [10, 15, 20] & 15 \\ \hline
Minimum samples split & [2, 5, 10] & 5 \\ \hline
Minimum leaf samples & [2, 4, 6] & 2 \\ \hline
\end{tabular}
\label{tab:parameters}
\end{center}
\vspace{-20pt}
\end{table}

\subsection{Model Inference} \label{5.3}
After training, CPGM can be used for inference. 
CPGM receives queries with metadata (e.g., data category and sensitivity level) and jurisdictions (e.g., source, target or both), and makes inference based on these data features. Finally, CPGM generates and returns the compliance policies.

\section{Implementation and Evaluation} \label{c6}
In this section, we implement a prototype system of CBCMS. 
We train, validate, and evaluate CPGM in Python.
We use Hyperledger Fabric \cite{fabric} as \emph{Blockchain Platform} of CBCMS. 
Based on this, we implement smart contracts supporting operations of users, metadata, and policies in Go.

Based on the implementation, we conduct comprehensive experiments to evaluate CBCMS's efficiency and effectiveness. 
We compare CPGM's inference accuracy with a rule-based baseline, further assessing its inference efficiency and concurrency performance. 
Additionally, we explore the performance of the blockchain as a pluggable component within CBCMS.

\subsection{Test Environment}
CPGM runs on a machine with 2 AMD EPYC 9554 CPUs (3.10GHz) and 756 GB RAM. 
We use Hyperledger Caliper \cite{caliper} to measure blockchain throughput and latency. 
The blockchain network operates across four physical machines, each with 16 vCPUs (Intel Xeon(R) Bronze 3206R, 1.9GHz) and 32 GB RAM. 
An additional machine with 8 vCPUs and 32 GB RAM is used to run Caliper, simulating client requests.

\begin{table*}[htbp]
    \centering
    \caption{Inference Accuracy Comparison of CPGM and Rule-Based Method on GDPR, CCPA and PIPL}
    \renewcommand{\arraystretch}{1.2} 
    \begin{tabular}{|c|c|c|ccc|ccc|c|}
    \hline
    \textbf{Category of Action} & \textbf{Specific Label} & \textbf{Support} & \multicolumn{3}{c|}{\textbf{CPGM}} & \multicolumn{3}{c|}{\textbf{Rule Based Method}} \\  
    \cline{4-6} \cline{7-9}  
    \textbf{Based on PDL} & & & \textbf{P(\%)} & \textbf{R(\%)} & \textbf{F1(\%)} & \textbf{P(\%)} & \textbf{R(\%)} & \textbf{F1(\%)} \\  
    \hline  
    & End-to-end Encryption & 428 & 97.24 & 98.83 & 98.03 & 94.44 & 63.55 & 75.98 \\  
    & Encrypt In Storage & 425 & 97.24 & 99.53 & 98.37 & 94.24 & 65.41 & 77.22 \\  
    & Pseudonymisation & 153 & 98.61 & 92.81 & 95.62 & 90.54 & 87.58 & 89.04 \\  
    GDPR & Anonymization & 77 & 98.48 & 84.42 & 90.91 & 41.33 & 80.52 & 64.63 \\
    Security Measures & Access Control & 428 & 97.93 & 99.53 & 98.73 & 94.50 & 64.25 & 76.50 \\
    & Availability and Recovery & 211 & 97.16 & 97.16 & 97.16 & 93.01 & 63.03 & 75.14 \\
    & Transparency in Processing & 74 & 96.97 & 86.49 & 91.43 & 97.28 & 50.53 & 66.51 \\
    & Regular Testing and Evaluation & 566 & 97.75 & 100 & 98.86 & 40.74 & 74.32 & 52.63 \\ \hline
    & Access & 209 & 96.68 & 97.61 & 97.14 & 91.95 & 65.55 & 76.54 \\
    GDPR & Rectification and Erasure & 207 & 94.31 & 96.14 & 95.22 & 90.28 & 62.80 & 74.07 \\
    Data Subject Rights & Portability & 147 & 95.74 & 91.84 & 93.75 & 43.90 & 85.71 & 58.06 \\
    & Restriction & 86 & 98.67 & 86.05 & 91.93 & 48.59 & 80.23 & 60.53 \\ \hline
    & Decision Based on Adequacy & 149 & 96.53 & 93.29 & 94.88 & 86.90 & 84.56 & 85.71 \\
    & Appropriate Safeguards & 150 & 98.61 & 94.67 & 96.60 & 91.37 & 84.67 & 87.89 \\
    & Exemptions in Certain Conditions & 150 & 97.92 & 94.00 & 95.92 & 91.61 & 87.33 & 89.42 \\
    & Lawfulness, Fairness and Transparency & 563 & 97.24 & 100 & 98.60 & 90.62 & 88.47 & 89.54 \\
    GDPR & Purpose Limitation & 428 & 97.70 & 99.30 & 98.49 & 97.32 & 51.69 & 67.52 \\
    Compliance Requirements& Data Minimization & 291 & 98.29 & 98.63 & 98.46 & 93.47 & 63.55 & 75.66 \\
    & Accuracy & 153 & 97.30 & 94.12 & 95.68 & 89.93 & 92.10 & 91.00 \\
    & Storage Limitation & 295 & 98.29 & 97.29 & 97.79 & 87.50 & 86.93 & 87.21 \\
    & Integrity and Confidentiality & 293 & 97.95 & 97.61 & 97.78 & 90.18 & 87.71 & 88.93 \\
    & Data Protection Officer & 285 & 96.23 & 98.60 & 97.40 & 85.16 & 92.63 & 88.74 \\ \hline
    GDPR & Cooperation with Authorities & 293 & 98.63 & 98.29 & 98.46 & 91.07 & 90.44 & 90.75 \\
    Supervision and Management& Management and Notification & 85 & 100 & 88.24 & 93.75 & 45.39 & 81.18 & 58.23 \\ \hline
    \multicolumn{2}{|c|}{\textbf{Summary of GDPR}} & \textbf{6146} & \textbf{97.56} & \textbf{95.18} & \textbf{96.29} & \textbf{81.72} & \textbf{76.45} & \textbf{76.56} \\ \hline
    & Access & 233 & 96.69 & 95.71 & 96.33 & 97.54 & 51.07 & 67.04 \\  
    & Deletion & 177 & 99.40 & 93.22 & 96.21 & 92.24 & 60.45 & 73.04 \\
    CCPA & Correction & 182 & 96.99 & 88.46 & 92.53 & 91.07 & 56.04 & 69.39 \\
    Data Subject Rights& Portability & 339 & 94.92 & 99.12 & 96.97 & 88.66 & 64.60 & 74.74 \\
    & Non-discrimination & 133 & 97.46 & 86.47 & 91.63 & 80.67 & 72.18 & 76.19 \\
    & Right to Know & 254 & 98.00 & 96.46 & 97.22 & 98.11 & 40.94 & 57.78 \\ \hline
    & Transparency and Responsibility & 232 & 96.96 & 96.12 & 96.54 & 98.37 & 52.16 & 68.17 \\
    CCPA & Disclosure of Processing & 468 & 97.50 & 100 & 98.73 & 97.69 & 54.27 & 69.78 \\
    Compliance Requirements& Data Minimization & 132 & 94.62 & 93.18 & 93.89 & 86.73 & 74.24 & 80.00 \\
    & Processing Limitations & 135 & 94.62 & 91.11 & 92.83 & 78.29 & 74.81 & 76.52 \\ \hline
    CCPA & Cooperation with Authorities & 354 & 98.31 & 98.31 & 98.31 & 89.45 & 64.69 & 75.08 \\
    Supervision and Management& Management and Notification & 70 & 96.67 & 82.86 & 89.23 & 40.87 & 67.14 & 50.81 \\ \hline
    \multicolumn{2}{|c|}{\textbf{Summary of CCPA}} & \textbf{2709} & \textbf{96.86} & \textbf{93.42} & \textbf{95.04} & \textbf{86.64} & \textbf{61.05} & \textbf{69.88} \\ \hline
    PIPL & Data Encryption & 412 & 98.10 & 100 & 99.04 & 97.99 & 47.33 & 63.83 \\
    Security Measures& Data De-identification & 294 & 97.32 & 98.98 & 98.15 & 89.00 & 63.27 & 73.96 \\
    & Data Anonymization & 105 & 96.15 & 95.24 & 95.69 & 78.90 & 81.90 & 80.37 \\ \hline
    & View and Copy & 155 & 97.99 & 94.19 & 96.05 & 91.67 & 56.77 & 70.12 \\
    PIPL & Rectification and Erasure & 149 & 95.97 & 95.97 & 95.97 & 87.76 & 57.72 & 69.64 \\
    Data Subject Rights& Portability & 200 & 98.46 & 96.00 & 97.22 & 85.56 & 80.00 & 82.69 \\
    & Right to Withdraw Consent & 95 & 97.80 & 93.68 & 95.70 & 85.42 & 86.32 & 85.86 \\ \hline
    & National Security Assessment & 106 & 98.08 & 96.23 & 97.14 & 83.65 & 82.08 & 82.86 \\
    & International Legal Requirements & 107 & 98.08 & 95.33 & 96.68 & 86.54 & 84.11 & 85.31 \\
    PIPL & International Cooperation & 114 & 97.30 & 94.74 & 96.00 & 49.00 & 85.96 & 62.42 \\
    Compliance Requirements& Internal Supervision & 199 & 98.97 & 96.98 & 97.97 & 82.09 & 82.91 & 82.50 \\
    & Data Protection Manager & 411 & 97.86 & 100 & 98.92 & 97.96 & 46.72 & 63.26 \\
    & Punishment for Illegal Acts & 303 & 99.33 & 98.02 & 98.67 & 90.59 & 60.40 & 72.48 \\ \hline
    PIPL & Cooperation with Authorities & 295 & 97.99 & 99.32 & 98.65 & 93.78 & 61.36 & 74.18 \\
    Supervision and Management& Management and Notification & 201 & 99.49 & 96.52 & 97.98 & 85.86 & 84.58 & 85.21 \\ \hline
    \multicolumn{2}{|c|}{\textbf{Summary of PIPL}} & \textbf{3146} & \textbf{97.93} & \textbf{96.75} & \textbf{97.32} & \textbf{85.72} & \textbf{70.76} & \textbf{75.65} \\ \hline
\end{tabular}  
\label{tab:performance_comparison}  
\end{table*} 


\begin{figure*}[htbp]
\centering
\begin{subfigure}{0.49\textwidth}
    \centering
    \includegraphics[width=0.9\textwidth]{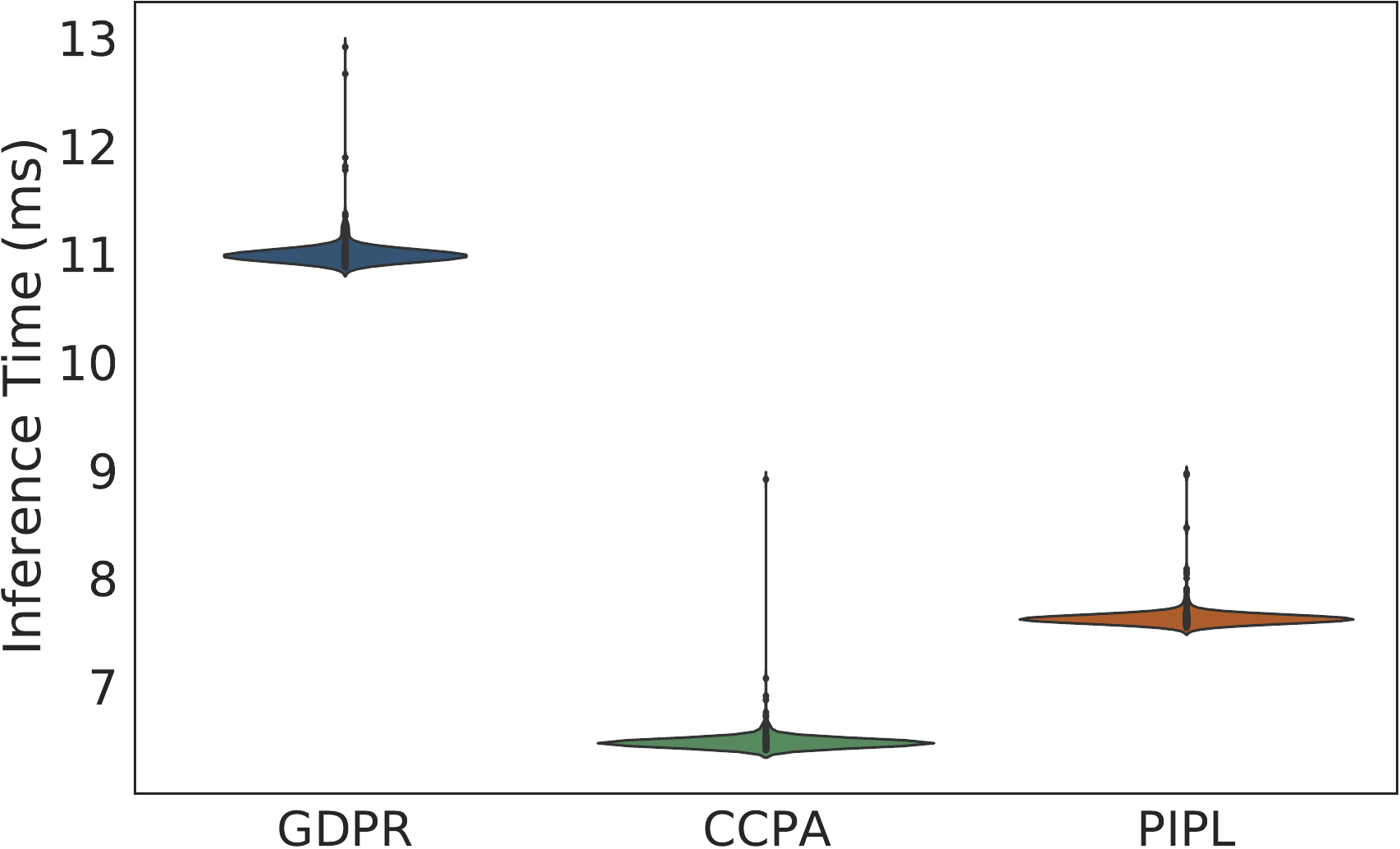}
    \caption{Inference time distribution of CPGM.}
    \label{fig:inf_distribution}
\end{subfigure}%
\hfill
\begin{subfigure}{0.49\textwidth}
    \centering
    \includegraphics[width=0.9\textwidth]{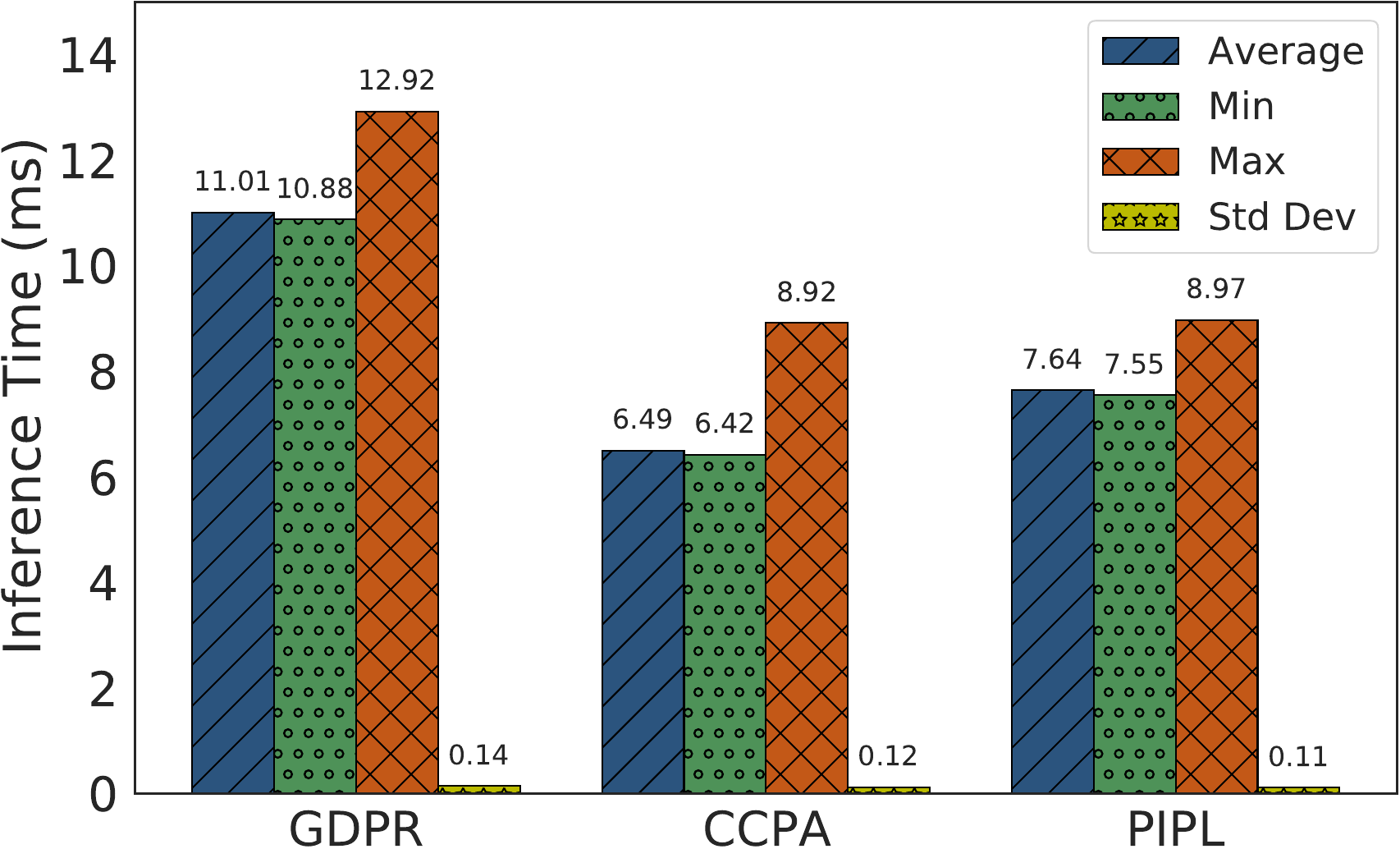}
    \caption{Inference time statistics of CPGM.}
    \label{fig:inf_time}
\end{subfigure}
\caption{Inference efficiency analysis of CPGM across different regulations (GDPR, CCPA, and PIPL).}
\label{fig:combined_inference_time}
\end{figure*}


\begin{figure*}[htbp]
\centering
\begin{subfigure}{0.49\textwidth}
    \centering
    \includegraphics[width=0.9\textwidth]{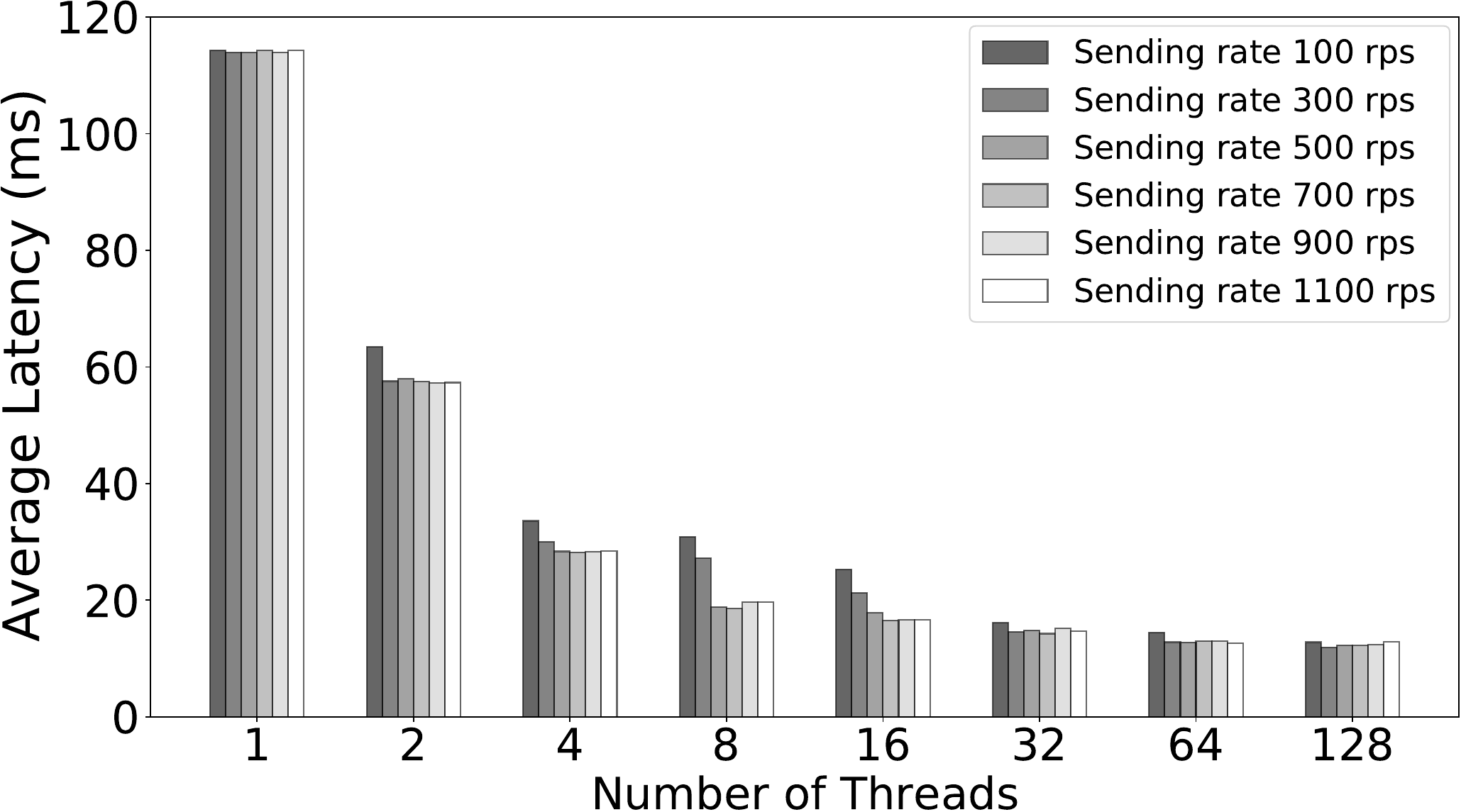}
    \caption{Average latency of CPGM under different workloads.}
    \label{fig:latency}
\end{subfigure}%
\hfill
\begin{subfigure}{0.49\textwidth}
    \centering
    \includegraphics[width=0.9\textwidth]{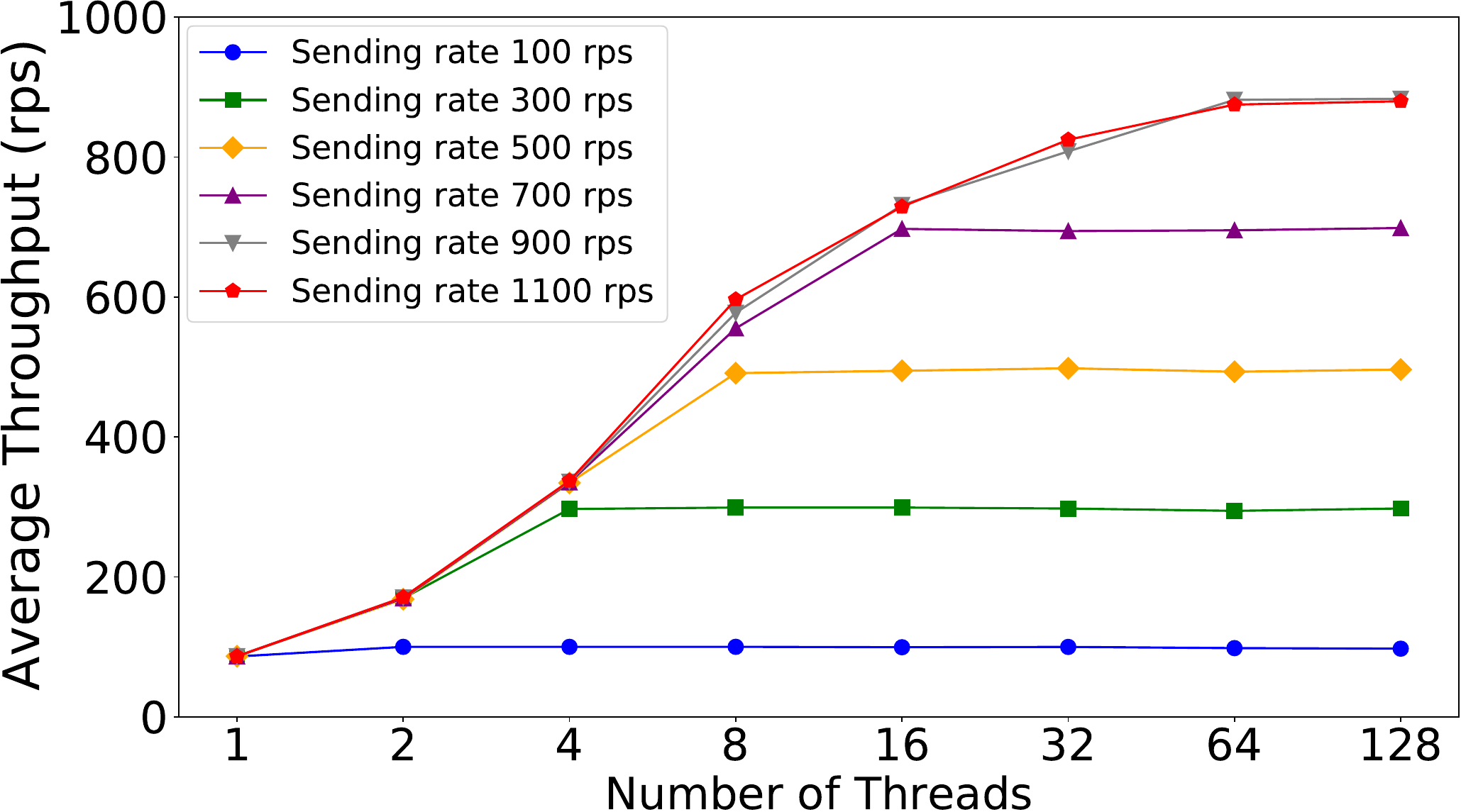}
    \caption{Average throughput of CPGM under different workloads.}
    \label{fig:throughput}
\end{subfigure}
\caption{Concurrency performance analysis of CPGM: Latency and throughput under various workloads.}
\label{fig:combined_performance}
\vspace{-5pt}
\end{figure*}

\subsection{Inference Accuracy of CPGM}
To fully demonstrate the performance of the CPGM model in CBCMS, we design a rule-based method as a baseline.
We define a set of fixed rules based on the sensitivity level and the jurisdiction role.
The specific implementation is as follows:

\begin{enumerate}
    \item \textbf{Initialization of Label Vectors}: Initially, all the values of the multi-label binary encoding are inactive (set the value to 0).
    
    \item \textbf{Policy Sets Based on Sensitivity Level}:
    \begin{itemize}
        \item \textbf{Non-sensitive data}: Activate (set the value to 1) basic data protection policies such as data encryption and transparency requirements.
        \item \textbf{Low-sensitive data}: Besides non-sensitive data policies, activate data subject rights protection policies, such as data access and rectification rights.
        \item \textbf{Medium-sensitive data}: Besides low-sensitive data policies, activate controls over cross-border data transfer and compliance requirements.
        \item \textbf{Highly-sensitive data}: Activate comprehensive data protection policies, including security measures, data subject rights protection, compliance requirements, and supervision actions.
    \end{itemize}
    This step results in a set of policies where more sensitive data leads to more policies being activated.

    \item \textbf{Policy Sets Based on Jurisdiction Role}:
    \begin{itemize}
        \item \textbf{Source Jurisdiction Role}: Activate policy labels specific to the source jurisdiction. 
        For example, if the source jurisdiction requires certain security measures, those labels will be activated.
        \item \textbf{Target Jurisdiction Role}: Activate policy labels specific to the target jurisdiction. 
        For example, if the target jurisdiction has stringent requirements for data subject rights, those labels will be activated.
    \end{itemize}
    This step results in a second set of policies, where the jurisdiction roles determine which policies are activated.

    \item \textbf{Final Label Determination}: The final label set is determined by performing an element-wise AND operation between the two sets generated in the previous steps. 
    Only the labels that are activated in both the sensitivity level and jurisdiction role sets will remain active (set to 1) in the final label vector.
\end{enumerate}

Table \ref{tab:performance_comparison} shows the inference accuracy of CPGM and the rule-based baseline on the same test dataset annotated and divided in Section \ref{5.1}.
Support represents the actual number of samples in each category, indicating the dataset's imbalance.
Thus, we use Macro Average to treat each category evenly.

CPGM shows superior performance across all metrics compared to the baseline. 
The highest precision achieved by CPGM is 97.93\%, with the greatest improvement over the baseline being 15.84\%. 
For recall, the maximum achieved by CPGM is 96.75\%, showing an enhancement of up to 32.37\% compared to the baseline. 
The F1 score, which balances precision and recall, reaches a maximum of 97.32\%, reflecting an improvement of up to 25.16\% over the baseline method.

Despite the baseline method's relatively good precision, stemming from its strict rules designed to avoid false positives, it struggles with recall due to its inability to adapt to complex data scenarios, leading to more false negatives. 
In contrast, CPGM excels by effectively considering multiple features, resulting in more accurate and balanced performance across diverse metrics.



\subsection{Inference Efficiency of CPGM}
To show the inference efficiency of the CPGM model in CBCMS, we measure its inference time under GDPR, CCPA, and PIPL, each measured 500 times for reliability.
Fig. \ref{fig:inf_distribution} illustrates the distribution of inference times using violin plots, while Fig. \ref{fig:inf_time} provides detailed statistics with a bar chart.

As shown in Fig. \ref{fig:inf_distribution} and Fig. \ref{fig:inf_time}, the inference times for CPGM are highly steady. 
The inference time for GDPR averages 11.01 ms, with a standard deviation of 0.14 ms. 
For CCPA, the average inference time is 6.49 ms with a standard deviation of 0.12 ms, while PIPL averages 7.64 ms with a standard deviation of 0.11 ms. 

These results highlight the minimal time cost and high stability of CPGM's inference process, ensuring that CBCMS meets real-time requirements for cross-border data transfer, supporting the dynamic demands of global data management.



\subsection{Concurrency Performance of CPGM}
To evaluate the concurrency performance of the CPGM model in CBCMS, we conduct tests with varying numbers of threads and request rates, as shown in Fig. \ref{fig:latency} and Fig. \ref{fig:throughput}.

Fig. \ref{fig:latency} indicates that increasing the number of threads, especially from 1 to 32 threads, significantly reduces the average latency.
When the thread counts are beyond 32, the latency stabilizes, demonstrating that the system efficiently utilizes multithreading to reduce processing time under high concurrency conditions.
Even at the highest request rate of 1100 rps, the system maintains a manageable latency of approximately 12-13 ms when using 128 threads, indicating robust scalability and efficient handling of high-load scenarios.

Fig. \ref{fig:throughput} indicates that the throughput shows a significant increase with the number of threads, especially up to 16 threads.
Beyond this point, the throughput continues to rise but at a diminishing rate, stabilizing around 64 threads.
Notably, the throughput remains consistently higher for higher request rates, indicating that the system can handle increased loads effectively without compromising performance.

The results show that CPGM effectively uses multithreading to manage high volumes of concurrent requests, significantly reducing latency and increasing throughput. 
This confirms CPGM's ability to maintain high performance under various load conditions, making CBCMS a strong solution for cross-border compliance management in demanding environments.

\subsection{Blockchain Performance}
\begin{figure}[tp]
\centering
\includegraphics[width=0.77\columnwidth]{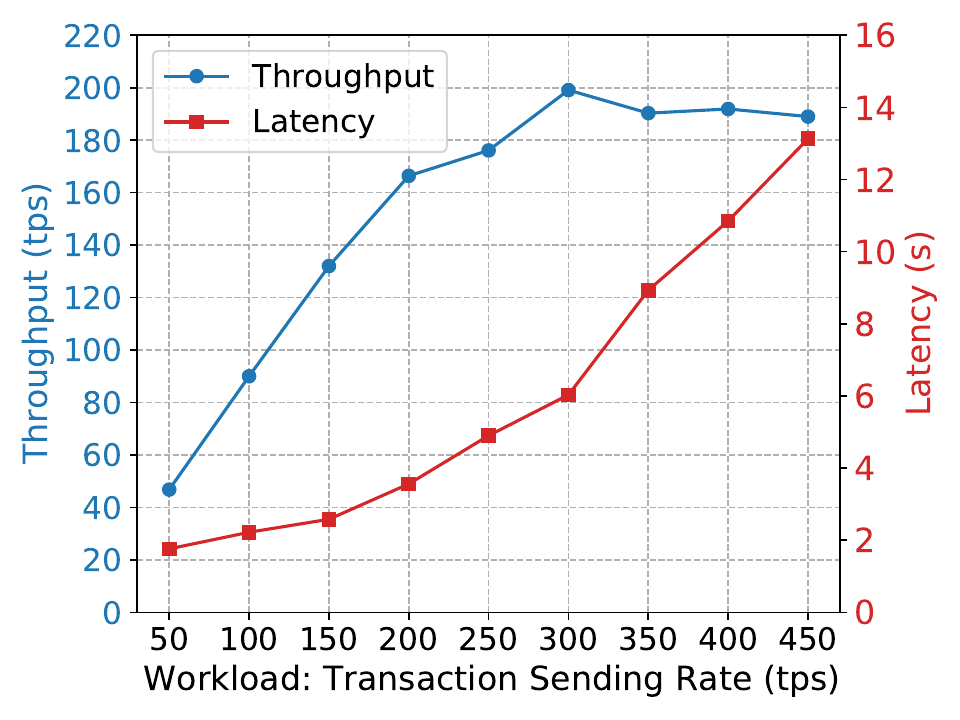}
\caption{Average throughput and latency of blockchain write operations under different workloads.}
\label{fig:blockchain_performance}
\vspace{-10pt}
\end{figure}

We conduct performance tests on the blockchain write operations in CBCMS using a setup of 5 orderer nodes and 5 peer nodes. 
The results show that throughput increases with the transaction rate, peaking at around 200-250 transactions per second (tps), with a maximum of approximately 200 tps. 
The average latency rises to about 13 seconds at 450 tps due to increased load, as shown in Fig. \ref{fig:blockchain_performance}.

The results highlight that while blockchain offers decentralized security, it also introduces performance trade-offs.  
However, blockchain is not essential for CBCMS, as the system can operate effectively without it. 
The primary advantage of incorporating blockchain lies in its ability to enhance security and trust through decentralization. 
Importantly, our tests demonstrate that the CBCMS remains efficient even with the added blockchain layer. 
Furthermore, CBCMS remains flexible and can integrate with various blockchain platforms by adjusting node configurations and consensus mechanisms, ensuring optimal performance across different platforms.

\section{Conclusion} \label{c7}

In this paper, we propose CBCMS to address compliance issues in cross-border data transfer.
CBCMS allows flexible configuration and extension of data processing policies, and supports unified policy management through PDL.
The CPGM model of CBCMS demonstrates high accuracy in generating compliance policies while ensuring high real-time and concurrent performance.
Our extensive experiments validate CBCMS's efficiency and effectiveness in compliance management across different jurisdictions.

\section*{Acknowledgment}
This research is funded by Fuxi Institution-CASICT Internet Infrastructure Laboratory.

\bibliographystyle{IEEEtran}
\bibliography{references}

\end{document}